\documentstyle[12pt]{article}
\begin{document}
\baselineskip 24pt
\newcommand{\mad}[1]{\frac{{\partial}_{\bar{z}}{#1}}{{\partial}_z
{#1}}}
\vspace{1.5cm}

\begin{center}
{\large \bf A CONFORMAL FIELD THEORY OF \\
\vspace{0.4cm}
 EXTRINSIC GEOMETRY OF 2-d SURFACES}
\end{center}

\vspace{2.0cm}

\begin{center}
K.S.Viswanathan{\footnote {e-mail address: kviswana@sfu.ca }} \\
Department of Physics \\
Simon Fraser University \\
Burnaby V5A 1S6, B.C., Canada. \\
and \\
R.Parthasarathy{\footnote {e-mail address: sarathy@imsc.ernet.in}} \\
Institute of Mathematical Sciences \\
Madras 600 113, India. \\
\end{center}

\vspace{5.0cm}

\noindent{Total number of pages: 30}
\newpage

\vspace{0.5cm}

\noindent{$\ \ $ Proposed Running Head}

\vspace{0.5cm}

\noindent{\bf $\ $ CONFORMAL FIELD THEORY OF EXTRINSIC GEOMETRY}

\vspace{2.5cm}

\noindent{Name and Mailing Address}

\vspace{0.5cm}

{\noindent{ K.S.Viswanathan \\
Department of Physics \\
Simon Fraser University \\
Burnaby B.C \\
Canada V5A 1S6 \\
\vspace{0.5cm}

Telephone: (604) 291 - 3157 \\
Fax: (604) 291 - 3592 \\}}
\newpage

\vspace{1.5cm}

\noindent{\it Abstract}

\vspace{0.5cm}
In the description of the extrinsic geometry of the string world sheet
regarded as a conformal immersion of a 2-d surface in $R^3$, it was
previously shown that, restricting to surfaces with $h\surd{g}\ =\ 1$,
where $h$ is the mean scalar curvature and $g$ is the determinant of the
induced metric on the surface, leads to Virasaro symmetry. An explicit
form of the effective action on such surfaces is constructed in this
article which is the extrinsic curvature analog of the WZNW action. This
action turns out to be the gauge invariant combination of the actions
encountered in 2-d intrinsic gravity theory in light-cone gauge and the
geometric action appearing in the quantization of the Virasaro group.
This action, besides exhibiting Virasaro symmetry in $z$-sector, has
$SL(2,C)$ conserved currents in the $\bar{z}$-sector. This allows us to
quantize this theory in the $\bar{z}$-sector along the lines of the WZNW
model. The quantum theory on $h\surd{g}\ =\ 1$ surfaces in $ R^3$ is
shown to be in the same universality class as the intrinsic 2-d gravity
theory.

\newpage
\noindent{\bf I.$\ \ $ INTRODUCTION}

\vspace{0.5cm}

Extrinsic geometry of surfaces immersed in $R^n$ plays an important role
in QCD-strings, 3-dimensional Ising model and in the study of biological
membranes. We have in our earlier publications [2,3] approached
this problem in terms of the Grassmannian $\sigma$-model.
In this approach, we make use of the Gauss map [4] of conformally immersed
surfaces in $R^n$ into the Grassmannian manifold $G_{2,n}$ realized as a
quadric in $CP^{n-1}$. This realization plays a key role in the Gauss map.
However, in order for the Grassmannian manifold to form tangent planes to
a given surface in $R^n$, $(n-2)$ integrability conditions must be
satisfied by the $G_{2,n}$ fields. These conditions have been explicitely
derived in [4] for $R^3$ and $R^4$ and by us [3] for $R^n$ ($n>4$). The
advantage of this approach to the string theory is that one is able to
rewrite the string action consisting of both the Nambu-Goto and extrinsic
curvature terms in terms of constrained K\"{a}hler $\sigma$ model action. In
a subsequent work [5] we have shown that in the geometry of surfaces
immersed in $R^3$ with the property $h\sqrt{g}\ =\ 1$, where $h$ is the
scalar mean curvature of and $g$ the determinant of the induced metric on
the surface, there is a hidden Virasaro symmetry. Specifically, it was
shown that $H_{zz}/\sqrt{g}$, where $H_{\alpha\beta}$ are the components
of the second fundamental form of the surface, transforms like energy
momentum tensor ($T_{zz}$) and that $H_{\bar{z}\bar{z}}$ transforms like
the metric tensor under the Virasaro transformation. It was further shown
that in $R^4$ one has analogous properties for surfaces with
$h\sqrt{g}\ =\ 1$, where $h\ =\ \sqrt(h^2_1 + h^2_2)$, and $h_1$ and $h_2$
are the two scalar mean curvatures along the two normals to the surface.
The hidden symmetry in $R^4$ is found to be Virasaro $\otimes $ Virasaro. A
simple way to see why one has doubling of the Virasaro algebra in $R^4$
is to observe that $G_{2,4}\simeq CP^1 \otimes CP^1$, while
$G_{2,3}\simeq CP^1$. For surfaces immersed in $R^n (n>4)$ not much is
known although it is expected that one would find W-gravities in certain
type of surfaces [6].

\vspace{0.5cm}

In this article, we continue our investigation of $h\sqrt{g}\ =\ 1$
surfaces in $R^3$. We construct an action on such surfaces involving
their extrinsic geometric properties. The procedure used here is a
variant of the one in [7,8], done in the context of 2-d intrinsic
gravity. In [8], Majorana fermions interacting with zweibein are
integrated out to obtain the {\it gravitational} WZNW action. Its
properties in the light-cone gauge (l.c) are unravelled in [7]. Here, we
consider 2-d fermions coupled however, to {\it extrinsic} curvature of
the surface on which the fermions live, by minimal coupling. The
minimal coupling for immersion in $R^3$, is through $SO(3)$
gauge fields determined by a local orthonormal frame on the surface.
Integrating out the fermions [9], we obtain the WZNW action which in this
case may be termed as {\it the induced extrinsic gravity action}. This
action has local $SO(3)$ gauge invariance. The restriction to $h\sqrt{g}\
=\ 1$ partially fixes the gauge for $A^a_z,\ (a=\pm,0)$ as the gauge
fields encode here the information about the extrinsic properties of the
immersed surface. Let us note that this restriction is not generally
covariant and clearly it is a gauge choice. It is equivalent to an algebraic
gauge condition. Our gauge choice is made on surfaces
realized in a conformal immersion. In
[10], the property of diffeomorphisms arising from partial gauge fixing
of $SL(2,R)$ gauge fields was considered. Here, we have found a {\it
physical system where these gauge fields and the partial gauge fixing
which were ad hoc in [10], arise naturally when treating $h\sqrt{g}\ =\
1$ surfaces}. The resulting action is given explicitly by
\begin{eqnarray}
{\Gamma}_{eff}(F_1,F_2) &=& {\Gamma}_{+}(F_2) + {\Gamma}_{-}(F_1)
-\frac{k}{4\pi}\int
\left(\frac{{\partial}_{\bar{z}}F_2}{{\partial}_zF_2}\right) D_zF_1 \
dz\wedge d\bar{z},
\end{eqnarray}
where,
\begin{eqnarray}
D_zF_1 &=& \frac{{\partial}^3_zF_1}{{\partial}_zF_1} -
\frac{3}{2}{\left(\frac{{\partial}^2_zF_1}{{\partial}_zF_1}\right)}^2
\end{eqnarray}
is the Schwarz derivative of $F_1$.

\vspace{0.5cm}

In (1), ${\Gamma}_{+}(F_2)$ is precisely the action that appears in
intrinsic 2-d gravity theory in l.c gauge [7], while ${\Gamma}_{-}(F_1)$
is the geometric action that one encounters in the quantization of the
Virasaro group by the method of coadjoint orbits [11]. The coupling term
in (1) is needed to make ${\Gamma}_{eff}(F_1,F_2)$ invariant under
Virasaro transformation of $F_1$ and $F_2$, which are related
to the second fundamental form as follows (see section II.);
\begin{eqnarray}
\frac{{\partial}_{\bar{z}}F_2}{{\partial}_zF_2} &=& H_{\bar{z}\bar{z}},
\end{eqnarray}
and
\begin{eqnarray}
D_zF_1 &=& H_{zz}/\sqrt{g}.
\end{eqnarray}
We find that (1) is invariant under Virasaro transformation, whose
infinitesimal version reads as
\begin{eqnarray}
{\delta}_{\epsilon(\bar{z},z)}(H_{zz}/\sqrt{g})&=&-\frac{k}{2}{\partial}
^3_z\epsilon(\bar{z},z)-2({\partial}_z\epsilon(\bar{z},z))H_{zz}/\sqrt
{g} \nonumber \\
&-&\epsilon(\bar{z},z){\partial}_z(H_{zz}/\sqrt{g}),
\end{eqnarray}
and
\begin{eqnarray}
{\delta}_{\epsilon(\bar{z},z)}(H_{\bar{z}\bar{z}})&=&\frac{k}{2}
{\partial}_{\bar{z}}\epsilon(\bar{z},z)+\epsilon(\bar{z},z){\partial}
_zH_{\bar{z}\bar{z}}
-({\partial}_z\epsilon)H_{\bar{z}\bar{z}}.
\end{eqnarray}
Transformations (5) and (6) are equivalent to the following changes in
$F_i$,
\begin{eqnarray}
{\delta}_{\epsilon}F_i &=& \epsilon(\bar{z},z){\partial}_zF_i;\ \ \
(i=1,2).
\end{eqnarray}
In what follows we shall call ${\Gamma}_{+}(F_2)$ as the l.c action,
while ${\Gamma}_{-}(F_1)$ as the geometric action. It should be
emphasized here (to avoid any confusion) that we are discussing a
2-d {\underline{extrinsic}} gravity theory (extrinsic curvature induced
2-d gravity theory) and $F_1$ and $F_2$ are related to extrinsic
geometry by (3) and (4), while the corresponding action in 2-d intrinsic
gravity theory is in terms of an $F$ which parameterizes the metric tensor
of the 2-d surface.

\vspace{0.5cm}

The equation of motion following from (1) is given by
\begin{eqnarray}
{\partial}^3_zH_{\bar{z}\bar{z}}&=&[{\partial}_{\bar{z}}-H_{\bar{z}\bar{z}}
{\partial}_z-2({\partial}_zH_{\bar{z}\bar{z}})]\frac{H_{zz}}{\sqrt{g}}.
\end{eqnarray}
This equation is identically satisfied if we take both the first
fundamental form (induced metric $g_{\alpha\beta}$) and the second
fundamental form $(H_{\alpha\beta})$ as determined by
$X^{\mu}(\bar{z},z)$. However, we will
see that it is convenient to regard $H_{zz}/\sqrt{g}$ and
$H_{\bar{z}\bar{z}}$ as independent fields, independent of $X^{\mu}$,
but are related by the equation of motion (8). It is further shown that
there is a useful composition formula for (1), which enables us to
write it either as l.c action for $F_2\circ F^{-1}_1\ =\
F_2(\bar{z},F^{-1}_1(\bar{z},z))$, where the inverse function is defined
through
\begin{eqnarray}
F_i(\bar{z},F^{-1}_i(\bar{z},z)) &\equiv& z,
\end{eqnarray}
or as the geometric action for $F_1\circ F^{-1}_2$. This important
correspondence enables us to find the conserved $SL(2,C)$ currents for
the action (1) in the $\bar{z}$ (left) sector of the theory. The total
energy momentum tensor $T^{tot}_{\bar{z}\bar{z}}$ is of the Sugawara
form. The theory of extrinsic geometry induced 2-d gravity action for
$h\sqrt{g}\ =\ 1$ surfaces in $R^3$ is then quantized in the left
sector. From these observations, it is clear that the quantum theory of
the left sector on $h\sqrt{g} \ =\ 1$ surfaces is in the same
universality class as the 2-d intrinsic gravity theory.

\vspace{0.5cm}

It is well known that the conventional extrinsic curvature action [1]
is given by
\begin{eqnarray}
S_P &=& \frac{1}{{{\alpha}_{0}}^2}\int {\mid H \mid}^2\sqrt{g} \
dz\wedge d\bar{z}.
\end{eqnarray}
It is interesting to note that the third term (interaction term) in (1)
is precisely (10). Thus we could say that (1) is the conformally invariant
extension of (10) where $H_{zz}/\sqrt{g}$ and $H_{\bar{z}\bar{z}}$ are
the dynamical fields.

\vspace{0.5cm}

It will be shown that $F_1$ and $F_2$ appearing in (1) are the
inhomogeneous coordinates of $G_{2,3}\simeq CP^1$ i.e., the Grassmannian
$\sigma$-model fields arising in the Gauss map. The paper is organized
as follows. In section II we review the basic notions of the Gauss map
of surfaces and the properties of $h\sqrt{g}\ =\ 1$ surfaces. In
section III, we derive the Virasaro invariant action for extrinsic
geometry induced 2-d gravity. The composition formulae and symmetries
of this action in both the left and right sectors are studied
in section IV. The theory is quantized in section V and the results are
summarized in section VI.

\vspace{0.5cm}

\noindent{\bf II.$\ \ $ BASIC PROPERTIES OF $h\sqrt{g}$ SURFACES IN $R^3$}

\vspace{0.5cm}

We shall briefly review the basic properties of extrinsic geometry of
surfaces in $R^3$ [see Ref.2,4 for details] in  the following
subsections.
\vspace{0.5cm}

(i) Let $X^{\mu}({\xi}^1,{\xi}^2)$, $(\mu \ =\ 1,2,3)$ be the immersion
coordinates of a surface $S$ in $R^3$. The induced metric on $S$ is
\begin{eqnarray}
g_{\alpha\beta} &=& {\partial}_{\alpha}X^{\mu} {\partial}_{\beta}X_{\mu}
\ \ \ \ (\alpha,\beta = 1,2).
\end{eqnarray}
The second fundamental form $H_{\alpha\beta}$ is defined by
\begin{eqnarray}
{\partial}_{\alpha}{\partial}_{\beta}X^{\mu}&=&{\Gamma}^{\gamma}_
{\alpha\beta}{\partial}_{\gamma}X^{\mu} + H_{\alpha\beta}N^{\mu},
\end{eqnarray}
and
\begin{eqnarray}
{\partial}_{\alpha}N^{\mu}&=&-H_{\alpha\beta}g^{\beta\gamma}{\partial}_
{\gamma}X^{\mu}.
\end{eqnarray}
${\Gamma}^{\gamma}_{\alpha\beta}$ are the affine connections determined
by the induced metric (11) and $N^{\mu}$ is the local normal to $S$. The
intrinsic scalar curvature $R$ is given by
\begin{eqnarray}
R&=& (H^{\alpha}_{\alpha})^2 - H^{\beta}_{\alpha} H^{\alpha}_{\beta}.
\end{eqnarray}
In 2-dimensions $\int R\sqrt{g}d^2\xi$ is a topological invariant.
However either of the two terms in (14) being {\it separately}
reparametrization invariant can be used as a Lagrangian density for
extrinsic curvature effects [1], modulo Euler characteristic.

\vspace{0.5cm}

Let us take $({\xi}^1,{\xi}^2)$ to be the isothermal coordinates on $S$.
We shall regard $S$ as a Riemann surface. Introducing complex
coordinates $z={\xi}^1 + i{\xi}^2$ and $\bar{z}={\xi}^1 -i{\xi}^2$,
the metric in a conformal gauge has the property
\begin{eqnarray}
g_{zz}&=& g_{\bar{z}\bar{z}}=0, \nonumber \\
g_{z\bar{z}}&=&g_{\bar{z}z}\ \neq\ 0.
\end{eqnarray}

\vspace{0.5cm}

(ii) The Grassmannian manifold $G_{2,n}$ is defined as a set of oriented
2-planes in $R^n$ passing through an origin. $G_{2,n}$ is
isomorphic to $SO(n)/SO(2)\times SO(n-2)$. This may also be represented
as a complex quadric  $Q_{n-2}$ in $CP^{n-1}$ defined by
\begin{eqnarray}
\sum^{n}_{k=1} {w_k}^2 &=& 0,
\end{eqnarray}
where $w_k \in C$ are homogeneous coordinates of $CP^{n-1}$. To see
this, we write $w_k\ =\ a_k + ib_k$, where $\{a_k\}$ and $\{b_k\}$ are
real. (16) now reads as
\begin{eqnarray}
\vec{A}\cdot \vec{B} &=& 0, \nonumber \\
\|\vec{A}\| &=& \|\vec{B}\|,
\end{eqnarray}
where $\vec{A}=(a_1,.....,a_n)$, $\vec{B}=(b_1,.....,b_n)$ and so
$\vec{A}$ and $\vec{B}$ form an orthogonal basis for an
oriented 2-plane in $R^n$.

\vspace{0.5cm}

(iii) The Gauss map $G:\ S\rightarrow G_{2,n}\simeq Q_{n-2}$ is defined
by
\begin{eqnarray}
{\partial}_z X^{\mu} &=& \Psi {\Phi}^{\mu};\ \ \ \ \ (\mu
=1,2,.......n),
\end{eqnarray}
where ${\Phi}^{\mu}$ is a point in $Q_{n-2}$ satisfying ${\Phi}^{\mu}\
{\Phi}^{\mu}\ =\ 0$ (see Eqn.(16)) and $\Psi (\bar{z},z)$ is yet to be
determined. Note that this mapping makes sense as the induced metric
$g_{\alpha\beta}={\partial}_{\alpha}X^{\mu}{\partial}_{\beta}X_{\mu}$,
satisfies (15) (conformal immersion). Note further that every immersion
in $R^n$ can be realized as a conformal immersion. While
${\partial}_{\bar{z}}{\partial}_zX^{\mu}$ is real, the term
${\partial}_{\bar{z}}(\Psi {\Phi}^{\mu})$ may not in general be real.
This together with the fact that ${\partial}_{\bar{z}}{\partial}_z
X^{\mu}$ is normal to $S$ (see Eqn.12) determine $\Psi$ and the
integrability conditions. From (12) using the notation $H^{\mu}\equiv
H^{\mu,\alpha}_{\alpha}\equiv H^{i\alpha}_{\alpha}N^{\mu}_i$,
(i=1,2,....n-2),
\begin{eqnarray}
H^{\mu} &=& \frac{2}{{\lambda}^2} {\partial}_{\bar{z}}{\partial}_z
X^{\mu},
\end{eqnarray}
where,
\begin{eqnarray}
{\lambda}^2 & = & 2{\| {\partial}_z X^{\mu} \| }^2 \
  =\  2{\mid \Psi \mid }^2 {\| \Phi \| }^2.
\end{eqnarray}
Expressing $H^{\mu}$ in terms of $\Psi$, ${\Phi}^{\mu}$ and noting
that $H^{\mu}$ is normal to the surface while ${\Phi}^{\mu}$ is
tangential, we obtain
\begin{eqnarray}
(\ell n \Psi)_{\bar{z}} &=& - \eta,
\end{eqnarray}
and
\begin{eqnarray}
V^{\mu} &=& {\partial}_{\bar{z}}{\Phi}^{\mu} - \eta {\Phi}^{\mu},
\nonumber \\
&=& \bar{\Psi}{\| \Phi \|}^2 H^{\mu},
\end{eqnarray}
where,
\begin{eqnarray}
\eta &=& ({\partial}_{\bar{z}}{\Phi}^{\mu}) {\bar{\Phi}}^{\mu}/{\| \Phi
\|}^2.
\end{eqnarray}
(21) and (22) together contain (n-2) integrability conditions
[2,3,4] on the Gauss map ${\Phi}^{\mu}$. What it means is that not every
$G_{2,n}$ field (${\Phi}^{\mu}$) forms a tangent plane to a given
surface. We notice
that the n-dimensional space-time emerges from the 2-dimensional
$G_{2,n} \ \sigma$-model, through the reconstruction of surfaces from the
$CP^1$ fields by
\begin{eqnarray}
X^{\mu}(z,\bar{z}) &=& \int^{z} Re\{\psi {\Phi}^{\mu}\ dz \}, \nonumber
\end{eqnarray}
where the integrals are path independent [4].

\vspace{0.5cm}

(iv) Let us work out the details of Gauss map in $R^3$. ${\Phi }^{\mu}$
may be expressed in terms of a single complex function $F(\bar{z},z)$
which is the inhomogeneous coordinate for $CP^1$ [2,4].
\begin{eqnarray}
{\Phi}^{\mu} &=& \left( 1-F^2, i(1+F^2), 2F \right).
\end{eqnarray}
The normal Gauss map $N^{\mu}$ is given by
\begin{eqnarray}
N^{\mu} &=& \frac{1}{(1+{\mid F\mid }^2)}\left(F+\bar{F}, -i(F-\bar{F}),
{\mid F\mid }^2-1 \right).
\end{eqnarray}
The Gauss map integrability condition is
\begin{eqnarray}
Im\  {\partial}_{\bar{z}} \left(
\frac{{\partial}_z{\partial}_{\bar{z}}F}{{\partial}_{\bar{z}}F}
-2\frac{\bar{F}{\partial}_zF}{1+{\mid F \mid }^2} \right) &=& 0.
\end{eqnarray}
This is a third order differential constraint. A straightforward
computation yields
\begin{eqnarray}
V^{\mu} &=& -2({\partial}_{\bar{z}}F) N^{\mu},
\end{eqnarray}
which allows us to write $\Psi $ in the form,
\begin{eqnarray}
\Psi (\bar{z},z) &=& -\frac{{\partial}_z\bar{F}}{h (1+{\mid F\mid }^2)^2},
\end{eqnarray}
where $h\ =\ H^{\mu}\ N^{\mu}$ is the scalar mean curvature.

\vspace{0.5cm}

(v) Let us introduce a local
orthonormal moving frame ${\hat{e}}_1,\ {\hat{e}}_2,\ {\hat{e}}_3\
\equiv \ N^{\mu}$, where ${\hat{e}}_1$ and ${\hat{e}}_2$ are tangential
to $S$. A convenient choice for ${\hat{e}}_i$ in terms of
its image in the Gauss map is as follows [5];
\begin{eqnarray}
{\hat{e}}_1&=&\frac{1}{(1+{\mid F\mid }^2)}\left(
1-\frac{F^2+{\bar{F}}^2}{2},\frac{i}{2}(F^2-{\bar{F}}^2),F+\bar{F}
\right),
\end{eqnarray}
\begin{eqnarray}
{\hat{e}}_2&=&\frac{1}{1+{\mid F\mid}^2}\left(
\frac{i}{2}(F^2-{\bar{F}}^2),1+\frac{F^2+{\bar{F}}^2}{2},-i(F-\bar{F})
\right),
\end{eqnarray}
\begin{eqnarray}
{\hat{e}}_3&=&\frac{1}{1+{\mid F\mid}^2}\left(
F+\bar{F},-i(F-\bar{F}),{\mid F\mid}^2-1\right).
\end{eqnarray}
The structure equations (12) and (13) may be written in the following
form:
\begin{eqnarray}
{\partial}_z{\hat{e}}_i &=& (A_z)_{ij}\ {\hat{e}}_j,
\end{eqnarray}
\begin{eqnarray}
{\partial}_{\bar{z}}{\hat{e}}_i &=& (A_{\bar{z}})_{ij}\ {\hat{e}}_j.
\end{eqnarray}
Using (29) to (31) one can express $A_z$ and $A_{\bar{z}}$ in terms of
$F(\bar{z},z)$ and its derivatives. From $(A_z)_{ij}\ =\
{\hat{e}}_j\ {\partial}_z{\hat{e}}_i$, we note that $(A_z)_{ij}\ ,\
(A_{\bar{z}})_{ij}$ are antisymmetric. Under a local gauge
transformation by ${\cal {G}} \in SO(3,C)$
\begin{eqnarray}
{\hat{e}}_i &\rightarrow & {\cal {G}}_{ij}\ {\hat{e}}_j,
\end{eqnarray}
a given surface $S$ is mapped into another surface $S^{\cal {G}}$. The
vector potential $A_{\mu}$ then transforms as a gauge field. Note
further that ${\hat{e}}_1$ and ${\hat{e}}_2$ are determined only upto an
$SO(2,C)$ rotation. In the basis (29) to (31), $(A_z)_{12}$ is
non-vanishing. However, it was shown in [5] that by rotating
${\hat{e}}_1,\ {\hat{e}}_2$ through an angle $\alpha$ given by
\begin{eqnarray}
\alpha &=& \frac{-i}{2} \ell n({\bar{\Psi}}^2\ {\| \Phi \|}^2),
\end{eqnarray}
where $\Psi$ and ${\Phi}^{\mu}$ were defined in (28), $A_z\ \rightarrow
\ {A'}_z$ such that $({A'}_z)_{12}\ =\ 0$. A straightforward computation
of the transformed gauge potentials yields the following (we drop the
prime on $A_z$ and $A_{\bar{z}}$ from now on),
\begin{eqnarray}
A_z&=&\left( \begin{array}{ccc}
0&0&\frac{1}{\surd{2}}(\frac{H_{zz}}{\surd{g}}+h\surd{g}) \\
 & & \\
0&0&\frac{i}{\surd{2}}(-\frac{H_{zz}}{\surd{g}}+h\surd{g}) \\
 & & \\
-\frac{1}{\surd{g}}(\frac{H_{zz}}{\surd{g}}+h\surd{g})&-\frac{i}
{\surd{2}}(-\frac{H_{zz}}{\surd{g}}+h\surd{g})&0 \\
\end{array}
\right),
\end{eqnarray}
and
\begin{eqnarray}
A_{\bar{z}}&=&\left( \begin{array}{ccc}
0&-[{\partial}_{\bar{z}}(\ell
n\bar{\Psi})+\frac{2F{\partial}_{\bar{z}}\bar{F}}{1+{\mid F\mid}^2}]&
\frac{1}{\surd{2}}(H_{\bar{z}\bar{z}}+h) \\
 & & \\
i[{\partial}_{\bar{z}}\ell n\bar{\Psi}+\frac{2F{\partial}_{\bar{z}}\bar
{F}}{1+{\mid F\mid}^2}]&0&\frac{i}{\surd{2}}(H_{\bar{z}\bar{z}}-h) \\
 & &  \\
-\frac{1}{\surd{2}}(H_{\bar{z}\bar{z}}+h)&-\frac{i}{\surd{2}}(H_{\bar
{z}\bar{z}}-h) &0 \\
\end{array}
\right).
\end{eqnarray}
In deriving (36) and (37) we have substituted back wherever possible, in
terms of the characteristics of the 2-d surface from the Gauss map.

We now restrict to surfaces for which
$h\surd{g}$ = 1 and investigate the residual gauge degrees of freedom
for $A_z$ which leaves it in the form (36) with only $H_{zz}/\surd{g}$
as unfixed. This residual symmetry is a diffeomorphism. Among the three
(complex) parameters determining $SO(3,C)$, two are determined in terms
of the third, by the restrictions $A^0_z\ =\ 0,\ A^-_z\ =\ 1$ ( here we
are using the notation $A_z=A^a_z T^a, a=0,\pm$, and $T^{\pm}$ are the
raising and lowering operators and $T^0$ is the Cartan generator for
$SO(3)$ ). If we compute the transformation properties of the remaining
degree of freedom in $A^a_z$, namely, $H_{zz}/\surd{g}$, we find
\begin{eqnarray}
{\delta}_{{\epsilon}^-(\bar{z},z)}(H_{zz}/\surd{g})&=&{\partial}^3_z
{\epsilon}^- + 2({\partial}_z{\epsilon}^-) (\frac{H_{zz}}{\surd{g}})
+ {\epsilon}^-{\partial}_z(\frac{H_{zz}}{\surd{g}}).
\end{eqnarray}
(38) is recognizable as the standard Virasaro action on energy
momentum tensor. Similarly, looking at the transformation of
$A^-_{\bar{z}}\equiv H_{\bar{z}\bar{z}}$, we find
\begin{eqnarray}
{\delta}_{{\epsilon}^-(\bar{z},z)}H_{\bar{z}\bar{z}}&=&
{\partial}_{\bar{z}}{\epsilon}^- +
{\epsilon}^-{\partial}_zH_{\bar{z}\bar{z}}
- ({\partial}_z{\epsilon}^-)H_{\bar{z}\bar{z}}.
\end{eqnarray}
We thus deduce that $H_{zz}/\surd{g}$ transforms like energy-momentum
tensor while $H_{\bar{z}\bar{z}}$ transforms like the metric tensor of a
2-d conformal field theory. We conclude that surfaces conformally
immersed in $R^3$ with $h\surd{g}\ =\ 1$, exhibit Virasaro symmetry.
This mechanism of generating diffeomorphism from gauge symmetry was
first considered by Polyakov [10] in a different context.

The consistency condition for (32) and (33) is found to be
\begin{eqnarray}
{\partial}_{\bar{z}} A_z - {\partial}_z A_{\bar{z}} + [A_z,A_{\bar{z}}]
&\equiv & F_{\bar{z}z} = 0.
\end{eqnarray}
(40) is identically satisfied if $g_{\alpha\beta}$ and
$H_{\alpha\beta}$ are constructed from $X^{\mu}(\bar{z},z)$ as in (11)
and (12). We will take the point of view that $H_{zz}/\surd{g}$ and
$H_{\bar{z}\bar{z}}$ may be taken as independent dynamical degrees of
freedom, independent of $X^{\mu}$ variables. Then the integrability
condition (40) which can be rewritten as
\begin{eqnarray}
{\partial}^3_zH_{\bar{z}\bar{z}}&=&\{ {\partial}_{\bar{z}} -
H_{\bar{z}\bar{z}}{\partial}_z
- 2({\partial}_z H_{\bar{z}\bar{z}}) \}(\frac{H_{zz}}{\surd{g}}),
\end{eqnarray}
will be taken as the equation of motion on $h\surd{g}\ =\ 1$ surfaces
in which $H_{zz}/\surd{g}$ and $H_{\bar{z}\bar{z}}$ will be treated as
independent degrees of freedom.

\vspace{0.5cm}

(vi) We derive useful parametrizations for $H_{\bar{z}\bar{z}}$ and
$H_{zz}/\surd{g}$ in terms of the Gauss map of $h\surd{g}\ =\ 1$
surfaces. These parametrizations are used extensively in the next
section. From (18), (24) and (28) we find that when
$h\surd{g}\ =\ 1$
\begin{eqnarray}
\Psi \ {\partial}_{\bar{z}}F &=& -\frac{1}{2}.
\end{eqnarray}
{}From (12) it follows that $H_{\alpha\beta}\ =\
N^{\mu}{\partial}_{\alpha}{\partial}_{\beta}X^{\mu}$. Using the Gauss
map, we find that $H_{\bar{z}\bar{z}}\ =\
-2\bar{\Psi}\ {\partial}_{\bar{z}}\bar{F}$. Combining this with (42) we
find
\begin{eqnarray}
H_{\bar{z}\bar{z}} &=&
\frac{{\partial}_{\bar{z}}\bar{F}}{{\partial}_z\bar{F}}.
\end{eqnarray}
For $h\sqrt{g}\ =\ 1$ surfaces it can be verified that the integrability
condition (26) is replaced by a stronger condition. Eliminating $\Psi$
from (21) and (42), we find that $F$ satisfies
\begin{eqnarray}
\frac{{\partial}_{\bar{z}}{\partial}_{\bar{z}}F}{{\partial}_{\bar{z}}F} -
\frac{2\bar{F}{\partial}_{\bar{z}}F}{1+{\mid F\mid}^2} &=& 0.
\end{eqnarray}
Using this, a straightforward computation yields the
following useful result for $H_{zz}/\surd{g}$ when $h\surd{g}\ =\ 1$;
\begin{eqnarray}
\frac{H_{zz}}{\surd{g}} &=&
\frac{{\partial}^3_z\bar{F}}{{\partial}_z\bar{F}} -
\frac{3}{2}{\left(
\frac{{\partial}^2_z\bar{F}}{{\partial}_z\bar{F}}\right)}^2,
\nonumber \\
&\equiv & D_z\bar{F}.
\end{eqnarray}
In deriving (43) and (45) in terms of their images in the Grassmannian
$G_{2,3}$, we have made use of the fact that both $H_{\bar{z}\bar{z}}$
and $H_{zz}/\surd{g}$ are determined by the immersion coordinate
$X^{\mu}$ of the surface $S$. In view of our remarks in (v) we take
these as independent dynamical fields and thus parametrize them in terms
of independent Gauss maps $F_1$ , $F_2$ as
\begin{eqnarray}
H_{\bar{z}\bar{z}} &=& \frac{{\partial}_{\bar{z}}F_2}{{\partial}_zF_2},
\end{eqnarray}
and
\begin{eqnarray}
\frac{H_{zz}}{\surd{g}} &=& D_zF_1.
\end{eqnarray}
We shall look for an effective action which depends on $F_1$ and $F_2$
related to $H_{zz}/\sqrt{g}$ and $H_{\bar{z}\bar{z}}$ through (46) and
(47). We shall use the equation of motion (41) to constrain these fields.

\vspace{0.5cm}

\noindent{\bf III.$\ \ $ INDUCED EXTRINSIC GRAVITY ACTION.}

\vspace{0.5cm}

In Section II, we showed that $h\surd{g}\ =\ 1$ surfaces exhibit Virasaro
symmetry. We derive an action on such surfaces that is invariant under
Virasaro transformations. The procedure used is as follows. We couple
2-d fermions in the vector representation of $SO(3)$ to the gauge
fields $A_z$, $A_{\bar{z}}$ discussed in the previous section
It is well known that upon integrating out the fermions,
one gets the WZNW action [9]. We next evaluate this action explicitly
in terms of $F_1$ and $F_2$ (see Equations 46,47) for the partially
gauge fixed $A_z$ given in (36), i.e., on $h\surd{g}\ =\ 1$ surfaces.
Such an action was derived earlier in Ref.10 in a different context. For
this reason, we shall be brief in our adaptation to
$h\surd{g}\ =\ 1$ surfaces.

\vspace{0.5cm}

The gauge invariant action ${\Gamma}_{eff}$ depends on $A_z$ and
$A_{\bar{z}}$ through the parametrizations $A_z\ =\ h^{-1}{\partial}_zh$
and $A_{\bar{z}} \ =\ g^{-1}{\partial}_{\bar{z}}g$; $h,g \in SO(3)$.
($h$ and $g$ here are not to be confused with mean curvature scalar and
determinant of the induced metric). Note that choosing $h$ and $g$ as
independent elements of the gauge group is consistent with taking
$H_{\bar{z}\bar{z}}$ and $H_{zz}/\surd{g}$ as independent of $X^{\mu}$.
Explicitly
\begin{eqnarray}
{\Gamma}_{eff}&=&{\Gamma}_{-}(A_z) + {\Gamma}_{+}(A_{\bar{z}}) -
\frac{k}{4\pi}Tr \int A_z \ A_{\bar{z}} \ dz\wedge d\bar{z},
\end{eqnarray}
where $k\ =\ n_f$ is the number of fermions and
\begin{eqnarray}
{\Gamma}_{-}(A_z) &=& \frac{k}{8\pi} Tr \int
[({\partial}_zh)h^{-1}({\partial}_{\bar{z}}h)h^{-1}]\  d^2\xi \nonumber
\\
&+& \frac{k}{12\pi} Tr \int
{\varepsilon}^{abc}(({\partial}_ah)h^{-1}({\partial}_bh)h^{-1}(
{\partial}_ch)h^{-1})\  d^2\xi dt,
\end{eqnarray}
and ${\Gamma}_{+}(A_{\bar{z}})$ is given by an expression similar to
(49) with $h$ replaced by $g$ and the sign of the WZ term changed. Gauge
invariance of ${\Gamma}_{eff}$ is ensured by the last term in (48).

\vspace{0.5cm}

Next we apply the gauge restriction to (49).
\begin{eqnarray}
A^{-}_z \equiv h\surd{g} = 1 &;& A^0_z = 0.
\end{eqnarray}
A straightforward calculation yields
\begin{eqnarray}
{\Gamma}_{-}(A^{-}_z
=1,A^{0}_z=0,A^{+}_z=H_{zz}/\surd{g}) &\equiv & S_{-}(F_1)
=\frac{k}{8\pi}\int
\left(\frac{{\partial}_{\bar{z}}F_1}{{\partial}_zF_1}\right) \nonumber \\
& &\left(\frac{
{\partial}^3_zF_1}{{\partial}_zF_1} - 2{\left(\frac{{\partial}^2_zF_1}
{{\partial}_zF_1}\right)}^2 \right) dz\wedge d\bar{z},
\end{eqnarray}
where we have used (47) to write $H_{zz}/\surd{g}$ in terms of $F_1$.
(51) is the geometric action for the Virasaro group studied in [11].
It is not so easy to find ${\Gamma}_{+}(A_{\bar{z}})$, when
$A_{\bar{z}}$ is given by (37). However, we can evaluate it explicitly
in a light-cone like gauge. Let us define the quantum action
$S_{+}(H_{\bar{z}\bar{z}})$ by
\begin{eqnarray}
exp(-S_{+}(H_{\bar{z}\bar{z}}))&=& \int [dA_{\bar{z}}]\delta
(A^{-}_{\bar{z}}-H_{\bar{z}\bar{z}})
 exp(-({\Gamma}_{+}(A_{\bar{z}})-\int
A^{+}_{\bar{z}})).
\end{eqnarray}
Evaluating (52) at the stationary path
\begin{eqnarray}
\frac{\delta {\Gamma}_{+}}{\delta {A^0_{\bar{z}}}}=J^0_z=0 &;&
\frac{\delta {\Gamma}_{+}}{\delta A^{+}_{\bar{z}}}=J^{-}_z=1,
\end{eqnarray}
we obtain the classical limit which can be explicitly evaluated
[10] as (Note that $H_{\bar{z}\bar{z}}$ is the component $A^{-}_{\bar{z}}$
in (37))
\begin{eqnarray}
S_{+}(F_2) &=&-\frac{k}{8\pi}\int
\frac{{\partial}^2_zF_2}{{\partial}_zF_2}\left(
\frac{{\partial}_z{\partial}_{\bar{z}}F_2}{{\partial}_zF_2}
- \frac{{\partial}^2_zF_2}{{\partial}_zF_2} \cdot
\frac{{\partial}_{\bar{z}}F_2}{{\partial}_zF_2}\right) dz\wedge
d\bar{z}.
\end{eqnarray}
(54) is the gauge fixed form of ${\Gamma}_{+}(A_{\bar{z}})$ in
the gauge determined by (53). Note that it is analogous to the
light-cone gauge fixing in 2-d intrinsic gravity theory. (54) is
precisely of the form of the light-cone action in 2-d intrinsic gravity
theory.

\vspace{0.5cm}

The variations of $S_{-}(F_1)$ and $S_{+}(F_2)$ under infinitesimal
changes (in $F_1$,$F_2$) are given by
\begin{eqnarray}
\delta S_{-} &=& -\frac{k}{4\pi} \int ({\delta}_{\epsilon}F_1)
\frac{{\partial}_{\bar{z}}(D_zF_1)}{{\partial}_zF_1}\  dz\wedge d\bar{z},
\end{eqnarray}
and
\begin{eqnarray}
\delta S_{+} &=& -\frac{k}{4\pi} \int ({\delta}_{\epsilon}F_2)
\frac{1}{{\partial}_zF_2} \cdot {\partial}^3_z
\left(\frac{{\partial}_{\bar{z}}F_2}{{\partial}_zF_2}\right)\  dz\wedge
d\bar{z}.
\end{eqnarray}
The third term on the right hand side of (48) in the gauge $A^0_z\ =\
0$, $A^{-}_z\ =\ 1$ becomes $\int (A^{+}_zA^{-}_{\bar{z}} +
A^{+}_{\bar{z}})d^2\xi$, of which the second term has been taken into
account in (52). The remaining term, i.e., $\int A^{+}_zA^{-}_{\bar{z}}$,
with the identification $A^{+}_z\ =\ H_{zz}/\surd{g}\ =\ D_zF_1$ and
$A^{-}_{\bar{z}}\ =\ H_{\bar{z}\bar{z}}\ =\
{\partial}_{\bar{z}}F_2/{\partial}_zF_2$ yields $\frac{k}{4\pi}\int
(D_zF_1)\left(\frac{{\partial}_{\bar{z}}F_2}{{\partial}_zF_2}\right)$
and so the total action on $h\surd{g}\ =\ 1$ surfaces reads
\begin{eqnarray}
{\Gamma}_{eff}(F_1,F_2)&=&\frac{k}{8\pi}\int
\frac{{\partial}_{\bar{z}}F_1}{{\partial}_zF_1}\left(
\frac{{\partial}^3_zF_1}{{\partial}_zF_1} -
2{\left(\frac{{\partial}^2_zF_1}{{\partial}_zF_1}\right)}^2\right) \ dz
\wedge d\bar{z} \nonumber \\
&-&\frac{k}{8\pi}\int \frac{{\partial}^2_zF_2}{{\partial}_zF_2}\left(
\frac{{\partial}_z{\partial}_{\bar{z}}F_2}{{\partial}_zF_2} -
\frac{{\partial}^2_zF_2}{{\partial}_zF_2}\cdot
\frac{{\partial}_{\bar{z}}F_2}{{\partial}_zF_2}\right)\ dz\wedge d\bar{z}
\nonumber \\
&-& \frac{k}{4\pi} \int \frac{{\partial}_{\bar{z}}F_2}{{\partial}_zF_2}
D_zF_1 \ dz\wedge d\bar{z}.
\end{eqnarray}
Equation (57) is our main result in this section We have derived the
extrinsic geometric gravitational WZNW action on $h\surd{g}\ =\ 1$
surfaces in the light-cone like gauge (53). The
induced extrinsic 2-d gravity action combines in a gauge invariant way
the geometric and light-cone actions which have been studied in the
context of 2-d intrinsic gravity. It is remarkable that a gauge
invariant interacting theory of these two actions has  physical
implication in the theory on $h\surd{g}\ =\ 1$ surfaces.

\vspace{0.5cm}

Note that restricted gauge transformations have yielded diffeomorphism
due to which $A^{+}_z$ which initially had conformal spin 1, has become
$H_{zz}/\surd{g}\ =\ D_zF_1$, a conformal spin 2 object. Same remark
applies to $A^{-}_{\bar{z}}\rightarrow
\frac{{\partial}_{\bar{z}}F_2}{{\partial}_zF_2}$, which acquires a
conformal spin 2. This suggests that $F_2$ should be regarded as a
primary conformal field with conformal weight $0$ in $\bar{z}$ sector.
Later, we show that quantum corrections modify this property of $F_2$.

\vspace{0.5cm}

\noindent{\bf IV. $\ \ $ SYMMETRY PROPERTIES OF ${\Gamma}_{eff}$.}

\vspace{0.5cm}

In this section we discuss the symmetry properties of the gauge fixed
action (57) on $h\surd{g}\ =\ 1$ surfaces in $R^3$. The equation of
motion for (57) is
\begin{eqnarray}
{\partial}^3_z \left(
\frac{{\partial}_{\bar{z}}F_2}{{\partial}_zF_2}\right) -
{\partial}_{\bar{z}}D_zF_1
-\left(\frac{{\partial}_{\bar{z}}F_2}{{\partial}_zF_2}\right)
{\partial}_zD_zF_1 -
2{\partial}_z\left(\frac{{\partial}_{\bar{z}}F_2}{{\partial}_zF_2}
\right)\  D_zF_1 &=& 0.
\end{eqnarray}
Using (46) and (47), it is readily seen that (58) is equivalent to (8)
which we referred to as the anomaly equation. It should be mentioned
here that varying (57) with respect to $F_1$ and $F_2$ independently
yielded the single equation of motion (58). Our objective is
to construct an effective action in terms of $H_{\bar{z}\bar{z}}$ and
$H_{zz}/\surd{g}$ through their parameterization (46) and (47) such that
these fields are independent of $X^{\mu}$.

\vspace{0.5cm}

Let us first verify that (57) is invariant under infinitesimal Virasaro
transformations (38) and (39). It can be easily verified that these
transformations for $H_{\bar{z}\bar{z}}$ and $H_{zz}/\surd{g}$ are
equivalent to the following transformation for $F_i$ (i=1,2).
\begin{eqnarray}
{\delta}_{{\epsilon}^{-}}F_i &=&{\epsilon}^{-}(\bar{z},z)\
{\partial}_zF_i, \ \ \ \ \ (i=1,2).
\end{eqnarray}
Invariance of (57) under (59) may be verified readily using (55) and
(56). Now (59) is the infinitesimal version of the finite
transformations;
\begin{eqnarray}
F_1(\bar{z},z)&\rightarrow & F_1(\bar{z},f(\bar{z},z)) \equiv F_1 \circ
f, \nonumber \\
F_2(\bar{z},z)&\rightarrow & F_2(\bar{z},f(\bar{z},z)) \equiv F_2 \circ
f.
\end{eqnarray}
The infinitesimal version (59) corresponds to
\begin{eqnarray}
f(\bar{z},z) &=& z + {\epsilon}^{-}(\bar{z},z).
\end{eqnarray}
We therefore conclude that
\begin{eqnarray}
{\Gamma}_{eff}(F_1,F_2) &=& {\Gamma}_{eff}(F_1 \circ f, F_2 \circ f).
\end{eqnarray}
Eqn.(62) has an interesting consequence. If we choose $f\ =\
F^{-1}_2(\bar{z},z)$, where the inverse function is defined by the
relation
\begin{eqnarray}
F_2(\bar{z},F^{-1}_2(\bar{z},z)) &=& z,
\end{eqnarray}
we find that
\begin{eqnarray}
{\Gamma}_{eff}(F_1,F_2) &=& {\Gamma}_{-}(F_1\circ F^{-1}_2)
\ =\  {\Gamma}_{+}(F_2\circ F^{-1}_1),
\end{eqnarray}
where the last equality follows by the interchange, $F_1 \leftrightarrow
 F_2$.
We thus
have derived the following useful composition formula.
\begin{eqnarray}
{\Gamma}_{+}(F_2\circ F^{-1}_1) &=& {\Gamma}_{-}(F_1\circ F^{-1}_2)
\nonumber \\
&=&{\Gamma}_{+}(F_2) + {\Gamma}_{-}(F_1) -\frac{k}{4\pi}\int
\frac{{\partial}_{\bar{z}}F_2}{{\partial}_zF_2}  D_zF_1.
\end{eqnarray}
{}From (64) we conclude that the properties of the induced extrinsic
gravity action can be understood from those of {\it either} the
light-cone action ${\Gamma}_{+}(F_2\circ F^{-1}_1)$ encountered in 2-d
(intrinsic) gravity theory with the replacement of $f$ of intrinsic 2-d
gravity theory by $F_2\circ F^{-1}_1$
{\it or} from those of ${\Gamma}_{-}(F_1\circ F^{-1}_2)$ i.e.,
the geometric action which arises in the quantization of the Virasaro
group by coadjoint orbits (with the replacement of $f$ by $F_1\circ
F^{-1}_2$).

\vspace{0.5cm}

The equation of motion for the light-cone action for $F_2\circ F^{-1}_1$
is [7]
\begin{eqnarray}
{\partial}^3_z \left( \frac{{\partial}_{\bar{z}}(F_2\circ
F^{-1}_1)}{{\partial}_z(F_2\circ F^{-1}_1)}\right) &=& 0.
\end{eqnarray}
By using the rules for derivatives of the {\it composed function }
$F_2\circ F^{-1}_1$ with respect to $z$ and $\bar{z}$, in terms of those
of $F_1$ and $F_2$, it is readily verified that (66) reduces to the
equation of motion (58). From (66), we conclude that
${\Gamma}_{eff}(F_1,F_2)$ has a hidden $SL(2,C)$ [at the algebra level
it is same as $SL(2,R)$] symmetry in the $\bar{z}$-sector as in the
intrinsic 2-d gravity theory. Recall that the $SL(2,C)$ currents
$J^a_{\bar{z}}(a\ =\ \pm,0)$ are defined through
\begin{eqnarray}
\frac{{\partial}_{\bar{z}}(F_2\circ F^{-1}_1)}{{\partial}_z(F_2\circ
F^{-1}_1)} &=& J^{(+)}_{\bar{z}}(\bar{z}) - 2zJ^{0}_{\bar{z}}
+ z^2 J^{-}_{\bar{z}}(\bar{z}),
\end{eqnarray}
where the currents are functions of $\bar{z}$ only. A straightforward
calculation yields the following expressions for the currents.
\begin{eqnarray}
J^{+}_{\bar{z}}&=&-\frac{k}{2}[({\partial}_zF_1)\left(
\frac{{\partial}_{\bar{z}}F_2}{{\partial}_zF_2}\right) -
{\partial}_{\bar{z}}F_1] \nonumber \\
&-&2F_1J^{0}_{\bar{z}} + F^2_1 J^{-}_{\bar{z}},
\end{eqnarray}
\begin{eqnarray}
J^{0}_{\bar{z}}&=& -\frac{k}{2} F_1J^{-}_{\bar{z}} +
\frac{k}{2}[\frac{{\partial}_z{\partial}_{\bar{z}}F_2}{{\partial}_zF_2}
- \left(\frac{{\partial}_{\bar{z}}F_2}{{\partial}_zF_2}\right)
\frac{{\partial}^2_zF_2}{{\partial}_zF_2} \nonumber \\
&-&\frac{{\partial}_z{\partial}_{\bar{z}}F_1}{{\partial}_zF_1} +
\frac{{\partial}_{\bar{z}}F_2}{{\partial}_zF_2}\cdot
\frac{{\partial}^2_zF_1}{{\partial}_zF_1}],
\end{eqnarray}
and
\begin{eqnarray}
J^{-}_{\bar{z}}&=&-\frac{k}{2}\left( \frac{{\partial}^2_zF_1\cdot
{\partial}_z{\partial}_{\bar{z}}F_1}{({\partial}_zF_1)^3} -
\frac{{\partial}_{\bar{z}}{\partial}^2_zF_1}{({\partial}_zF_1)^2}\right)
\nonumber \\
&-&-\frac{k}{2({\partial}_zF_1)}[{\partial}^2_z\left(\frac{{\partial}_
{\bar{z}}F_2}{{\partial}_zF_2}\right) + \left(\frac{{\partial}^2_zF_1}
{{\partial}_zF_1}\right) {\partial}_z\left(\frac{{\partial}_{\bar{z}}
F_2}{{\partial}_zF_2}\right) \nonumber \\
&+& \frac{{\partial}_{\bar{z}}F_2}{{\partial}_zF_2} [\frac{{\partial}
^3_zF_1}{{\partial}_zF_1} - \frac{({\partial}^2_zF_1)^2}{({\partial}_zF_
1)^2}]].
\end{eqnarray}
It is straightforward to verify that
\begin{eqnarray}
{\partial}_zJ^{a}_{\bar{z}} &=& 0.
\end{eqnarray}
In order to exhibit $SL(2,C)$ invariance of the action (57), it is
convenient to identify it with the geometric action for $F_1\circ
F^{-1}_2$. Recall that [11] the geometric action is invariant under
$SL(2,C)$ transformations:
\begin{eqnarray}
F_1\circ F^{-1}_2 &\stackrel{SL(2,C)}{\rightarrow }& \frac{a(\bar{z})
F_1\circ F^{-1}_2 + b(\bar{z})}{c(\bar{z}) F_1\circ F^{-1}_2 +
d(\bar{z})},
\end{eqnarray}
with $ad\ -\ bc\ =\ 1$. Noether currents for these transformations,
whose infinitesimal form we write as
\begin{eqnarray}
{\delta}_{SL(2,C)} F_1\circ F^{-1}_2 &=& {\epsilon}_{-}(\bar{z}) +
{\epsilon}_{0}(\bar{z}) F_1\circ F^{-1}_2
+ {\epsilon}_{+}(\bar{z}) (F_1\circ F^{-1}_2)^2,
\end{eqnarray}
are precisely those given in (68) to (70).

\vspace{0.5cm}

Next, we discuss the consequences of reparametrization invariance of
(57). In the $\bar{z}$-sector, an infinitesimal conformal
transformation results in
\begin{eqnarray}
{\delta}_{\bar{\epsilon}}F_i &=&
\bar{\epsilon}(\bar{z})\ {\partial}_{\bar{z}}F_i;\ \ \ \ \ (i=1,2).
\end{eqnarray}
Expressing the change in ${\Gamma}_{eff}$ as
\begin{eqnarray}
{\delta}_{\bar{\epsilon}}{\Gamma}_{eff}(F_1,F_2) &=& -\frac{1}{2\pi}
\int dz\wedge d\bar{z}\
\bar{\epsilon}(\bar{z})\ {\partial}_zT_{\bar{z}\bar{z}},
\end{eqnarray}
we find
\begin{eqnarray}
T_{\bar{z}\bar{z}}(F_1,F_2) &=& T_{\bar{z}\bar{z}}(F_1) +
T_{\bar{z}\bar{z}}(F_2) + T^{Int}_{\bar{z}\bar{z}}(F_1,F_2),
\end{eqnarray}
where,
\begin{eqnarray}
T_{\bar{z}\bar{z}}(F_1) &=&
-\frac{k}{2}\{ \frac{({\partial}_{\bar{z}}F_1)({\partial}^2_z{\partial}_
{\bar{z}}F_1)}{({\partial}_zF_1)^2} -
\frac{1}{2}{\left(\frac{{\partial}_{\bar{z}}{\partial}_zF_1}{{\partial}
_zF_1}\right)}^2
- \frac{{\partial}^2_zF_1 {\partial}_{\bar{z}}F_1 {\partial}_z
{\partial}_{\bar{z}}F_1}{({\partial}_zF_1)^2}\},
\end{eqnarray}
\begin{eqnarray}
T_{\bar{z}\bar{z}}(F_2)&=&\frac{k}{2}[{\left({\partial}_z\left(\frac
{{\partial}_{\bar{z}}F_2}{{\partial}_zF_2}\right)\right)}^2-2\left(
\frac{{\partial}_{\bar{z}}F_2}{{\partial}_zF_2}\right){\partial}^2_z
\left(\frac{{\partial}_{\bar{z}}F_2}{{\partial}_zF_2}\right)],
\end{eqnarray}
and
\begin{eqnarray}
T^{Int}_{\bar{z}\bar{z}}(F_1,F_2) &=& \frac{k}{2}
\{\left(\mad{F_1}\right)[{\partial}_z\left(\mad{F_2}\right)
\frac{{\partial}^2_zF_1}{{\partial}_zF_1} \nonumber \\
&+&\left(\mad{F_2}\right)\left(\frac{{\partial}^3_zF_1}
{{\partial}_zF_1} - {\left(\frac{{\partial}^2_zF_1}
{{\partial}_zF_1}\right)}^2\right) + {\partial}^2_z
\left(\mad{F_2}\right)] \nonumber \\
&-&\frac{{\partial}_z{\partial}_{\bar{z}}F_1}{{\partial}
_zF_1}[2\mad{F_2}\left(\frac{{\partial}^2_zF_1}{{\partial}
_zF_1}\right) + {\partial}_z\left(\mad{F_2}\right)] \nonumber
\\
&+&
\frac{{\partial}^2_z{\partial}_{\bar{z}}F_1}{{\partial}_zF_1}
\left(\mad{F_2}\right) - {\left(\mad{F_2}\right)}^2 D_zF_1\}
\end{eqnarray}
(77) and (78) can be recognized to be the corresponding
$T_{\bar{z}\bar{z}}$ for the geometric and light-cone actions
respectively.

\vspace{0.5cm}

It is readily verified that, using the equation of motion (58),
\begin{eqnarray}
{\partial}_z T_{\bar{z}\bar{z}}(F_1,F_2) &=& 0.
\end{eqnarray}
We expect $T_{\bar{z}\bar{z}}(F_1,F_2)$ to be of Sugawara form, and
indeed, it can be verified that
\begin{eqnarray}
T_{\bar{z}\bar{z}}(F_1,F_2) &=& -\frac{1}{k} {\eta}_{ab} J^a_{\bar{z}}
J^b_{\bar{z}},
\end{eqnarray}
where,
\begin{eqnarray}
{\eta}_{00} = 1;\  & &{\eta}_{1-1}= {\eta}_{-11} = -\frac{1}{2}.
\end{eqnarray}
We thus conclude that the induced 2-d extrinsic gravity theory on
$h\surd{g}\ =\ 1$ surfaces in $R^3$ has conformal invariance in the
$\bar{z}$-sector. Furthermore, this conformal field theory is in the
same universality class as the induced 2-d intrinsic gravity theory.

\vspace{0.5cm}

Let us briefly discuss the energy-momentum tensor $T_{zz}(F_1,F_2)$.
Under infinitesimal conformal change
\begin{eqnarray}
{\delta}_{\epsilon (z)}F_i(\bar{z},z) &=& \epsilon (z)\
{\partial}_zF_i(\bar{z},z);\ \ \ \ (i=1,2).
\end{eqnarray}
This leads to
\begin{eqnarray}
{\delta}_{\epsilon }{\Gamma}_{eff}(F_1,F_2) &=& \frac{1}{2\pi} \int
\epsilon (z){\bigtriangledown}_{\bar{z}}(F_2)T_{zz},
\end{eqnarray}
where,
\begin{eqnarray}
{\bigtriangledown}_{\bar{z}}(F_2)T_{zz} &=& {\partial}_{\bar{z}}T_{zz} -
\left(\mad{F_2}\right)
{\partial}_zT_{zz}-2{\partial}_z\left(\mad{F_2}\right)T_{zz},
\end{eqnarray}
and
\begin{eqnarray}
T_{zz}(F_1,F_2) &=& \frac{k}{2}[D_zF_1 - D_zF_2].
\end{eqnarray}
{}From the equation of motion, one finds that
\begin{eqnarray}
{\bigtriangledown}_{\bar{z}}(F_2)T_{zz}(F_1,F_2) &=& 0.
\end{eqnarray}
The transformation of the $SL(2,C)$ currents and
$T_{\bar{z}\bar{z}}(F_1,F_2)$ under conformal and \\
$SL(2,C)$ transformations can be easily worked out. We give below the
results for completeness and will be used in quantizing the theory in
the next section

\vspace{0.5cm}

\noindent{\it{(i) $SL(2,C)$ Variations:}}

\vspace{0.5cm}

\begin{eqnarray}
{\delta}_{SL(2,C)} J_{\bar{z}} &=&
-\frac{k}{2}({\partial}_{\bar{z}}\epsilon) - [\epsilon, J_{\bar{z}}],
\end{eqnarray}
where,
\begin{eqnarray}
J_{\bar{z}} &=& {\eta}_{ab} T^a J^a_{\bar{z}},
\end{eqnarray}
\begin{eqnarray}
\epsilon &=& {\epsilon}_a(\bar{z}) T^a,
\end{eqnarray}
\begin{eqnarray}
T^0 = \frac{1}{2} \left[ \begin{array}{cc}
-1 & 0 \\
 0 & 1 \\
\end{array}\right], T^{+}&=&\left[ \begin{array}{cc}
0&0 \\
1&0 \\
\end{array}\right], T^{-} = \left[ \begin{array}{cc}
0 & -1 \\
0 & 0 \\
\end{array}\right]
\end{eqnarray}
\begin{eqnarray}
[T^a,T^b] &=& f^{ab}_c T^c,
\end{eqnarray}
\begin{eqnarray}
f^{1-1}_0 = -2, f^{01}_1 &=& -f^{0-1}_{-1} = 1.
\end{eqnarray}

\vspace{0.5cm}

\noindent{\it{(ii) Conformal Variations:}}

\begin{eqnarray}
{\delta}_{\bar{\epsilon}(\bar{z})}J^a_{\bar{z}}&=&\bar{\epsilon}
(\bar{z}) {\partial}_zJ^a_{\bar{z}}+({\partial}_{\bar{z}}\bar
{\epsilon})J^a_{\bar{z}},\ \ (a=\pm,0),
\end{eqnarray}
and
\begin{eqnarray}
{\delta}_{\bar{\epsilon}(\bar{z})}T_{\bar{z}\bar{z}}(F_1,F_2)=
\bar{\epsilon}(\bar{z}){\partial}_{\bar{z}}T_{\bar{z}\bar{z}}
(F_1,F_2)+2({\partial}_{\bar{z}}\bar{\epsilon})T_{\bar{z}
\bar{z}}(F_1,F_2).
\end{eqnarray}
(88) and (94) follow from (68)-(70) and (80) using
\begin{eqnarray}
{\delta}_{\bar{\epsilon}(\bar{z})}F_i &=&
\bar{\epsilon}(\bar{z}){\partial}_{\bar{z}}F_i,\ \ \ \ (i=1,2).
\end{eqnarray}

\vspace{0.5cm}

Before quantizing the induced extrinsic gravity theory in the
$\bar{z}$-sector, we must first modify the conformal properties of $F_1$.
If we regard ${\Gamma}_{eff}$ in (57) as equivalent to the geometric
action for $F_1\cdot F^{-1}_2\ \equiv \
F_1(\bar{z},F^{-1}_2(\bar{z},z))$, then following Aoyoma [12], one
should modify the conformal weight of $F_1\cdot F^{-1}_2$ in the
$\bar{z}$-sector, such that
\begin{eqnarray}
{\delta}_{\bar{\epsilon}(\bar{z})}(F_1\cdot F^{-1}_2) &=&
\bar{\epsilon}(\bar{z}){\partial}_{\bar{z}}(F_1\cdot F^{-1}_2) +
({\partial}_{\bar{z}}\bar{\epsilon})(F_1\cdot F^{-1}_2).
\end{eqnarray}
(97) which assigns conformal spin 1 for
$F_1$ in $\bar{z}$-sector, is necessitated by the equivalence of the
geometric action with the Liouville theory.

\vspace{0.5cm}

The same conclusion can also be arrived at looking from the point of
view of the light-cone action. Following [7], let us observe that (57)
expressed as ${\Gamma}_{+}(F_2\cdot F^{-1}_1)$ is the gauge fixed form
of the action on $h\surd{g}\ =\ 1$ surfaces, in the gauge,
$\frac{\delta {\Gamma}_{+}}{\delta A^0_{\bar{z}}}=0$ ,$
\frac{\delta{\Gamma}_{+}}{\delta A^{+}_{\bar{z}}}=1$ (See (53)). The
quantum action $S_{+}(F_2\cdot F^{-1}_1)$ which guarantees general
covariance will contain a term proportional to $\int \left(\frac{\delta
{\Gamma}_{+}}{\delta A^{-}_{\bar{z}}}\right) T_{\bar{z}\bar{z}}$. This
has been shown in [7] to yield an extra term to $T_{\bar{z}\bar{z}}$
which is ${\partial}_{\bar{z}}J^0_{\bar{z}}$. This is equivalent to the
modification given in terms of $F_1$ and $F_2$ as
\begin{eqnarray}
{\delta}_{\bar{\epsilon}(\bar{z})}F_1 &=&
\bar{\epsilon}(\bar{z}){\partial}_{\bar{z}}F_1 +
({\partial}_{\bar{z}}\bar{\epsilon})F_1, \nonumber \\
{\delta}_{\bar{\epsilon}(\bar{z})}F_2 &=&
\bar{\epsilon}(\bar{z}){\partial}_{\bar{z}}F_2.
\end{eqnarray}
Under this modification, (80) still holds with the modified
energy-momentum tensor
\begin{eqnarray}
T^{mod}_{\bar{z}\bar{z}}(\bar{z}) &=& -\frac{1}{k} {\eta}^{ab}
J^a_{\bar{z}} J^b_{\bar{z}} - {\partial}_{\bar{z}}J^0_{\bar{z}}.
\end{eqnarray}
Under the modified conformal transformations, the conformal weight of
the currents also change. We find instead of (94)
\begin{eqnarray}
{\delta}_{\bar{\epsilon}(\bar{z})}J^a_{\bar{z}} &=&
\left(\bar{\epsilon}(\bar{z}) {\partial}_{\bar{z}} +
(1+a){\partial}_{\bar{z}}\bar{\epsilon}(\bar{z})\right) J^a_{\bar{z}},
\end{eqnarray}
while, (95) remains valid for $T^{mod}_{\bar{z}\bar{z}}(F_1,F_2)$.

\vspace{0.5cm}

\noindent{\bf V.$\ \ $ QUANTUM THEORY ON $h\surd{g}\ =\ 1$ SURFACES IMMERSED
IN $R^3$}

\vspace{0.5cm}

We have all the machinery in place for quantizing our theory on
$h\surd{g}\ =\ 1$
surfaces in $R^3$. The $\bar{z}$-sector of the action (57) can be
quantized like the WZNW action [13]. The details of this procedure are
hardly new and we are contend with merely summarizing the essential
steps. From (68) to (70), we can expand $J^a_{\bar{z}}$ in Laurent
series as
\begin{eqnarray}
J^{a}_{\bar{z}} &=& \sum_{n=-\infty}^{\infty} J^{a}_n {\bar{z}}^{-n-1-a},
\end{eqnarray}
while from (99), $T_{\bar{z}\bar{z}}^{mod}$ is expanded as
\begin{eqnarray}
T^{mod}_{\bar{z}\bar{z}} &=& \sum_{n=-\infty}^{\infty} L_n
{\bar{z}}^{-n-2}.
\end{eqnarray}
Then, consistent with the $SL(2,C)$ transformations (88), the currents
are quantized as [see also [14]]
\begin{eqnarray}
[J^{a}_{n}, J^{b}_{m}] &=&-\frac{k}{2}(n+a){\eta}^{ab}{\delta}_{n+m,0} +
f^{ab}_{ c} J^c_{n+m},
\end{eqnarray}
with
\begin{eqnarray}
J^a_n \mid 0> &=& 0,
\end{eqnarray}
for $n\ >\ -a$. Next, we define the quantum version of the
energy-momentum tensor $T_{\bar{z}\bar{z}}$ or $L_n$. As shown
in [13], it is given by
\begin{eqnarray}
L_n &=& \frac{1}{(k+2)} \sum_{m=-\infty}^{\infty}{\eta}_{ab} :J^a_m
J^b_{n-m}: + (n+1)J^0_n,
\end{eqnarray}
where the normal ordering $(:\ :)$ is understood as
\begin{eqnarray}
\sum_{m=-\infty}^{\infty}:J^a_m J^b_{n-m}: &=& \sum_{m=-\infty}^{(n+b)}
J^a_m J^b_{n-m}
+ \sum_{(m=n+b+1)}^{\infty} J^b_{n-m} J^a_m.
\end{eqnarray}
We are now in a position to calculate the rest of the commutators. We
find [12]
\begin{eqnarray}
[L_n,J^{a}_m] &=& -(m-an) J^a_{n+m} - \frac{k}{2} n(n+1){\eta}^{a0}
{\delta}_{n+m},
\end{eqnarray}
and
\begin{eqnarray}
[L_n,L_m] &=& (n-m)L_{n+m} + \frac{c_L}{12} n(n-1)(n+1){\delta}_{n+m,0},
\end{eqnarray}
where,
\begin{eqnarray}
c_L &=& \frac{3k}{k+2} - 6k, \nonumber \\
&=& 15 - 6(k+2) - \frac{6}{k+2}.
\end{eqnarray}
Note that $-6k$ comes from the modified term in $T_{\bar{z}\bar{z}}$.
Thus the central charge of the Virasaro algebra in the left sector on
$h\surd{g}\ =\ 1$ surfaces is the {\it same} as for the 2-d intrinsic
gravity in the same sector.

\vspace{0.5cm}

It is interesting to note that in Section II, we observed that
conformally immersed surfaces in $R^3$  with $h\surd{g}\ =\ 1$
exhibit Virasaro invariance with $\frac{H_{zz}}{\surd{g}}\ \equiv
D_zF_1$ transforming like the energy-momentum tensor and yet, it is the
$\bar{z}$-sector of this theory which can be successfully quantized. For
the $z$-sector because of (87), the central charge remains unknown.

\vspace{0.5cm}

\noindent{\bf $\ \ $ VI.CONCLUSIONS}

\vspace{0.5cm}

In this article we have studied the classical and quantum properties of
a class of conformally immersed surfaces in background $R^3$. The
class of surfaces is characterized by their extrinsic geometric
property; i.e., $h\surd{g}\ =\ 1$. We have constructed an action on such
surfaces which depends on $\frac{H_{zz}}{\surd{g}}$ and
$H_{\bar{z}\bar{z}}$ as dynamical field variables. Here $H_{zz}$ and
$H_{\bar{z}\bar{z}}$ are components of the second fundamental form of
the surface and $g$ is the determinant of the induced metric on the
surface. The construction of the extrinsic geometric WZNW action is done
by minimally coupling 2-d fermions to gauge fields on the surface,
constructed from the structure equations to the surface. We
show that the gauge restriction  $h\surd{g}\ =\ 1$
has an explicit representation which is equivalent to a
gauge invariant coupling of light-cone and geometric type actions
studied in the literature in connection with intrinsic 2-d gravity
theory.

\vspace{0.5cm}

We emphasize in this article the importance of the role of the Gauss map
in establishing the existence of Virasaro symmetry in $h\surd{g}\ =\ 1$
surfaces. The Gauss map of conformally immeresed surfaces in $R^n$,
studied in detail elsewhere, is the map from a surface in the
Grassmannian manifold $G_{2,n}$. Since, not every Grassmannian manifold
field forms tangent planes to a surface in $R^n$, $(n-2)$
integrability conditions must be satisfied by the $G_{2,n}$ fields.
The effective action ${\Gamma}_{eff}(F_1,F_2)$ is a function of two
complex functions $F_1$, and $F_2$ which parameterize $H_{zz}/
\surd{g}\ =\ D_zF_1$, and $H_{\bar{z}\bar{z}}\ =\ \mad{F_2}$ and are
shown to be the image of $X^{\mu}$ in $G_{2,3}$. The effective action
${\Gamma}_{eff}(F_1,F_2)\
=\ {\Gamma}_{-}(F_1\cdot F^{-1}_2)\ =\ {\Gamma}_{+}(F_2\cdot F^{-1}_1)$
is shown to be the Grassmannian $\sigma$-model action for induced
extrinsic curvature action. Because of a useful composition formula for
${\Gamma}_{eff}(F_1,F_2)$, it is found that the induced 2-d extrinsic
gravity theory in $R^3$ on $h\surd{g}\ =\ 1$ surfaces is in the
same universality class as the intrinsic 2-d gravity theory. This result
constitutes the main observation in this work. Quantization of
$h\surd{g}\ =\ 1$ surfaces is carried out with the $SL(2,C)$ currents in
the $\bar{z}$-sector of the theory.

\vspace{0.5cm}

\noindent{\bf Acknowledgements}

We are thankful to I.Volovich and I.Y.Arefeva for useful discussions.
This work has been supported by an operating grant from the Natural
Sciences and Engineering Research Council of Canada. One of the authors
(R.P) thanks the Department of Physics for the kind hospitality during
his stay at Simon Fraser University where part of the work was done.

\vspace{0.5cm}

\noindent{\bf References}

\begin{enumerate}

\item A.M.Polyakov, Nucl.Phys. {\bf B268} (1986) 406; \\
      W.Helfrich, J.Phys.(Paris) {\bf 46} (1985) 1263; \\
      L.Peliti and S.Leibler, Phys.Rev.Lett. {\bf 54} (1985) 1690.

\item K.S.Viswanathan, R.Parthasarathy and D.Kay, Ann.Phys.(N.Y) \\
      (1991) 237.

\item R.Parthasarathy and K.S.Viswanathan, Int.J.Mod.Phys. {\bf A7} \\
      (1992) 1819.

\item D.A.Hoffman and R.Osserman, Proc.London.Math.Soc.(3) {\bf 50} \\
      (1985) 21; J.Diff.Geom. {\bf 18} (1983) 733.

\item R.Parthasarathy and K.S.Viswanathan, Int.J.Mod.Phys. {\bf A7} \\
      (1992) 317.

\item G.M.Sotkov, M.Stanishkov and C.J.Zhu, Nucl.Phys. {\bf B356}  \\
      (1991) 245.

\item A.M.Polyakov, Mod.Phys.Lett. {\bf A2} (1987) 893.

\item V.G.Knizhnik, A.M.Polyakov and A.B.Zamalodchikov, Mod.Phys. \\
      Lett. {\bf A3} (1988) 819.

\item A.M.Polyakov, in {\it Fields, Strings and Critical Phenomena} \\
      Les Houches, 1988; Eds.E.Brezin and J.Zinn-Justin. \\
      North Holland (1990).

\item A.M.Polyakov, Int.J.Mod.Phys. {\bf A5} (1990) 833. (see also \\
      in {\it Physics and Mathematics of Strings}; Eds.L.Brink, \\
      D.Friedman and A.M.Polyakov: World Scientific;1990,Singapore.)

\item A.Alekseev and S.Shatashvili, Nucl.Phys. {\bf B323} (1989) \\
      719.

\item S.Aoyoma, Int.J.Mod.Phys. {\bf A7} (1992) 5761.

\item V.G.Knizhnik and A.B.Zamolodchikov, Nucl.Phys. {\bf B247} \\
      (1984) 83.

\item M.E.Peskin, SLAC Pub.4251, 1987.
\end{enumerate}
\end{document}